# Gamma-ray bursts and black holes. Interpretation of the observational data.

A. V. Kuznetsov

Space Research Institute, 117810, Profsouznaya 84/32, GSP-7, Moscow, Russia

**Abstract.** The paper relies upon (Kuznetsov 2000), where the cosmic gamma - ray bursts (GRBs) in clusters are shown to have an apparent shift relative to the position of the source. Hence the main conclusion follows: only cluster of GRBs ensures correct localization of the source whereas determining the direction of arrival of a single GRB practically carries no information about the position of its source. Generation of repeated GRBs in the activity phase of the source, which amounts to 3-4 months and with the period of repetition of activity equal to 4 years are most general properties of clusters. These properties precisely correspond to the characteristics of manifestation of activity of X-ray NOVAE (BHXN), thus pointing that the source of GRB clusters considered is a black hole (BH) in the XN. The direct identification of the GRB source with the BHXN GRO J0422+32 presented here this confirms. The investigated properties of GRB clusters permit a final solution of 2 main problems of astrophysics: search for black holes and origin of cosmic GRBs.

## 1. Introduction

Since 1973 cosmic gamma-ray bursts have been most intensively studied over about 30 years. In the initial period of these studies it seemed natural to assume a Galactic origin of the bursts, but it has never been experimentally confirmed. A large-scale anisotropy expected in the case of Galactic origin has not been discovered. The distribution of GRBs appeared highly isotropic (Briggs et al. 1996) pointing to a possibility of cosmological origin of gamma-ray bursts.

The accompanying electromagnetic emission was first detected for GRB 970228 in the optics (van Paradijs et al. 1997) and in X-rays (Costa et al. 1997). For GRB 970508 afterglow was observed in the radio wavelengths (Frail & Kulkarni 1997) and in the optical, the latter permitted the redshift $z = 0.835$ to be determined (Metzger et al. 1997). If we assume that the measured redshift value is associated with the Doppler effect then the distance to the GRB source amounts to ~ Gpc. As of today, the $z$-value is determined for ~ 20 GRBs it ranges from 0.43 to 4.5.

The accuracy of GRB localization from the optical afterglow reaches several angular seconds, this, however, at best ensures identification with a host galaxy, of the lowest possible stellar magnitude, up to ~ 28. Despite the obvious success in studying this phenomenon, the GRB source as well as its mechanism remains unknown.

Searching for sources of repeated GRBs proved to be negative. It was shown that there is no point concentration of GRBs (Meegan et al. 1995; Tegmark et al. 1996). At the same time in 1996-1997 papers appeared (Kuznetsov 1996, 1997) about the discoveries of clustering of GRBs, which seem unusual because of their apparent extent: on the average it was about 20 degrees of arc. The true position of a GRB source coincides with the center of a GRB cluster. It is assumed that the observed phenomenon is an image of a source of repeated GRBs. The logical conclusion, which then follows from the existence of clusters – images of GRB sources, lies in the fact that the observed direction of GRB arrival is their apparent direction. It does not coincide with the position of the source. In this case localization of a single GRB, however accurate it might be, practically carries no information about the position of its source. By the existing concept: one GRB – one source, the coordinates of a GRB and its source would differ by about 10-12 degrees of arc. As a result, clusters – since they do exist - should be used for localizing GRB sources. An important characteristic of GRB clusters is that they are recurrent. Among 94 identified sources of GRBs twenty three times appear twice and one was observed three times. As was estimated during 9 years of *BATSE* observations, the interval between the repeated manifestations of a source activity ranges from 1 to 5 years, the phase of activity lasts from 1 to 12 months. On the average, the period of repetition is about 4 years and the duration of an active phase is 3-4 months. Comparison of these characteristics with those analogous for X-ray NOVAE shows their very good coincidence. It may be said that GRB clusters discover BH X-ray NOVAE as sources of GRBs and the obtained here identification this confirms. The observed image of the GRB source evidently may be explained on the base of the local space-time curvature formed by a black hole.

Fig.1 shows a cluster of 7 GRBs from the catalog (Kuznetsov 2000), observed during 970503-970820 with the center coordinates l, b = 142°.1 +34°.2. The position of a source coincides with the cluster center (black circle). The circles show the localization errors with the exception of a source. This cluster includes GRB970503, GRB970507, GRB970508, GRB970612.3, GRB970713.3, GRB970803 and GRB970820. The GRB970508 (diamond) is the well known burst from whose optical afterglow the redshift $z = 0.835$ was measured for the first time.

The catalog of X-ray binary systems (van Paradijs 1995), a list of 24 candidates to black holes in the low-mass binary systems (White & van Paradijs 1996) and 309 gamma-ray sources (Macomb & Gehrels 1999) were mainly used for possible identification of GRB sources. Two types of objects are singled out which may be identified with GRB sources. These are representatives of X-ray NOVAE GRO J0422+32 and GS 1354-645, which contain black holes as compact objects, as well as nearest galaxies M31, M33, M51, M81. The identification of the GRB source with GS 1354-645 is determined by the following gamma-ray bursts: 980103.1, 980124.3, 980126, 980218.3, 980310.2, 980325.1 and 980401.3. A catalog of about 100 GRB clusters discovered during 1991-2000 is given (Kuznetsov 2000). Only identification of GRO J0422+32 is considered here.

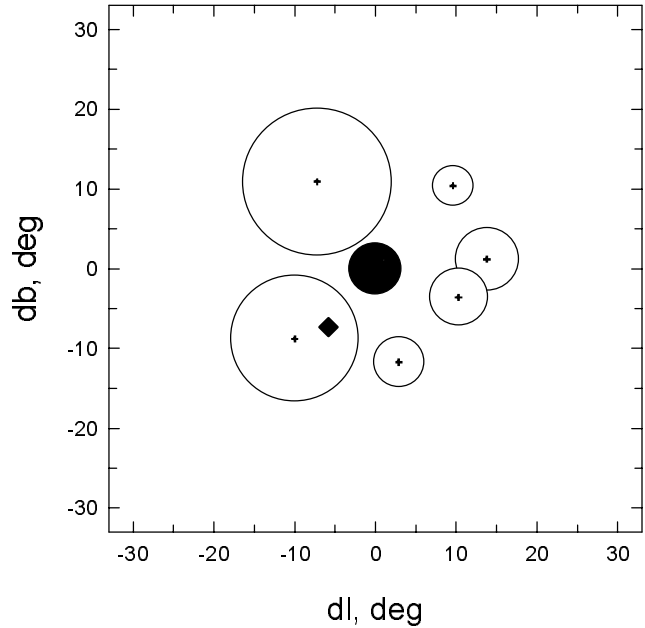

**Fig. 1**. The GRB cluster – the image of a source

## 2. Identification of the GRB sources with BH X-ray NOVAE

Some typical features discovered during direct identification of a source of repeated cosmic GRB clusters with a black hole of GRO J0422+32 are given below.

**Table 1. Two identified clusters of gamma-ray bursts**

1) l, b = 167.°9 - 5.°7

| GRB | TJD | $I$, erg cm$^{-2}$ s$^{-1}$ | $T_{90}$, s | $Q$, erg | $L$, erg s$^{-1}$ |
|---|---|---|---|---|---|
| 921110.3 | 8936 | 5.82E-08 | 41.1 | 3.14E+34 | 7.63E+32 |
| 921112.4 | 8938 | 1.23E-07 | 15.4 | 2.49E+34 | 1.62E+33 |
| 930104.1 | 8991 | 9.62E-07 | 0.6 | 7.57E+33 | 1.26E+34 |
| 930109 | 8996 | ? | (40) | ? | ? |
| 930123 | 9010 | 1.93E-07 | 22.3 | 5.66E+34 | 2.54E+33 |

2) l, b = 165.°4 - 9.°6

| GRB | TJD | $I$, erg cm$^{-2}$ s$^{-1}$ | $T_{90}$, s | $Q$, erg | $L$, erg s$^{-1}$ |
|---|---|---|---|---|---|
| 930926 | 9256 | 6.38E-08 | 89.4 | 7.48E+34 | 8.37E+32 |
| 931016.2 | 9276 | 3.00E-08 | 56.0 | 2.20E+34 | 3.93E+32 |
| 931115 | 9306 | 4.41E-08 | 31.3 | 1.81E+34 | 5.78E+32 |
| 931211 | 9332 | 1.56E-06 | 0.2 | 4.08E+33 | 2.04E+34 |
| 931229.2 | 9350 | ? | ? | ? | ? |
| 940222 | 9405 | 3.35E-07 | 11.1 | 4.88E+33 | 4.40E+32 |

Note. The intensity ($I$), the total energy in a burst ($Q$) and the luminosity ($L$) have been calculated using the data (GRB duration $T_{90}$ and fluence $F > 20$ keV) from the *BATSE* catalog (Paciesas et al. 1999). TJD – truncated Julian day (TJD = JD – 2,440,000). For calculations it is assumed that distance $D = 2.4$ kpc and the GRB radiation angle $\vartheta = 1$ deg.; l, b – coordinates of the cluster center.

Table 1 offers two clusters from the catalog (Kuznetsov 2000) the centers of which coincide in their position with the source GRO J0422+32.

The energy emitted in a GRB (fluence -$F$) and duration $T_{90}$ have a strong correlation, which determines the luminosity (intensity) curve with maximum in the GRB clusters, see Fig. 2.

Fig. 2 illustrates profiles of the light curves of GRO J0422+32 in X-ray (20 – 100 keV) and optical ranges, taken from (Chen et al. 1997; Callanan et al. 1995) as well as gamma-ray bursts of two above mentioned clusters. In each cluster, points corresponding to the intensities (ergs cm$^{-2}$ s$^{-1}$) of GRBs form curves with peaks. Vertical dotted lines determine time axes of bursts whose intensities are not known.

The first cluster of GRBs in the Figure has a delay of about 140 days relative to the main outburst and coincides in time with the secondary maximum (glitch) in the X-ray light curve. There is no emission peak (Callanan et al. 1995) in the $R$-band of the optical curve corresponding to the glitch, though the existence of the weak excess might be assumed. So Figures (Callanan et al. 1995; Goranski et al. 1996) demonstrate possible increase in the emission intensity for a given time instant, but the significance of this value was not estimated. The light curve in the radio wavelengths here shows a maximum of radiation (Shrader et al. 1994). As is seen from Fig. 2, the picture of the first cluster is not complete because of the lack of GRBs, which may be explained by a miscalculation of the events.

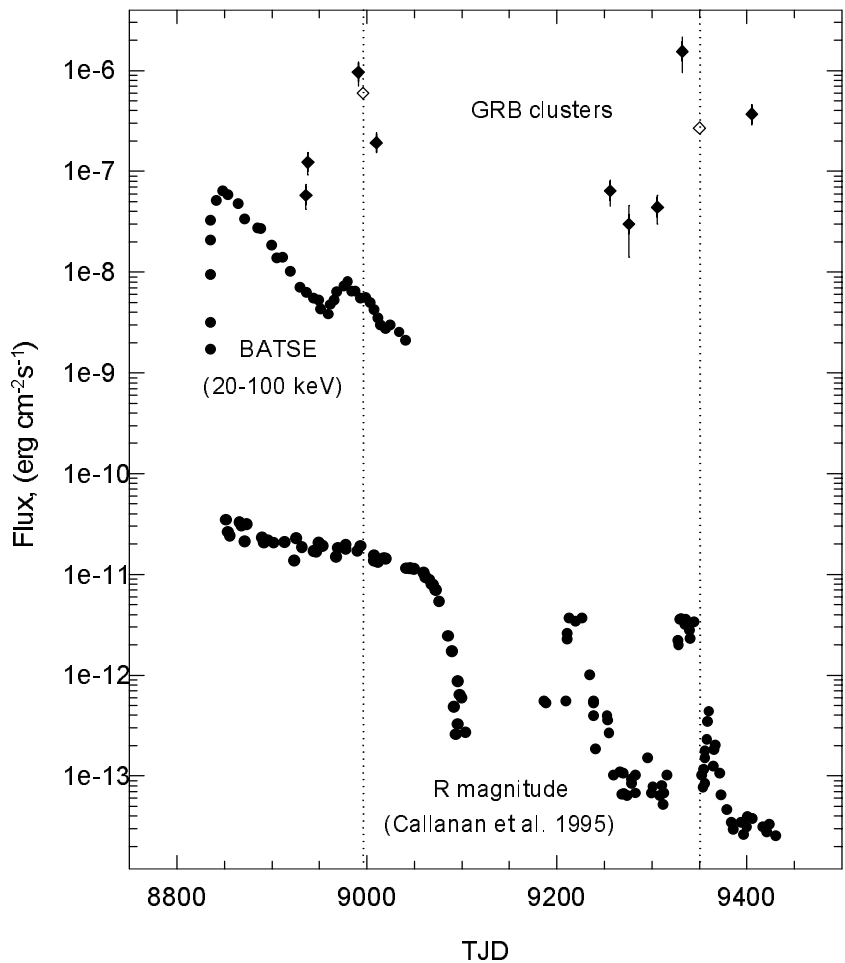

**Fig. 2.** GRB clusters and the light curves of GRO J0422 + 32

The second cluster surprisingly well copies a second minioutburst profile in the optical that makes it possible to assume with confidence the position of the 5th GRB (an empty diamond) in the given cluster. The profile formed by GRBs and minioutburst on the light curve in the optical coincide both in shape and amplitude which is determined as a maximum to minimum ratio and is ~ 50. The 6-th GRB coincides in time with the maximum of the weak last minioutburst in the optics. It may be an important sign that GRBs strongly correlate with minioutbursts on the light curves independently of minioutburst amplitude.

GRB intensities in two clusters both maximal and minimal are close in their values, though it seems that the first cluster is associated with a pulse on the X-ray curve, whereas the second, with the optical minioutburst. If this is the case, it points to the standard mechanism of GRB clusters generation. The reason of lack of GRBs corresponding to the first optical minioutburst is unknown (it may be a miscalculation of events) though the end of its decay is followed by 2 GRBs, and the source GRO J0422+32 was at the same time in the *BATSE* field of view (Shrader et al. 1997).

It is of interest that if we had at our disposal only the light curve in X-rays the identification of GRB source would not be so evident. Fig. 2 implies that the light curve of XN in the optical has a better compatibility with GRB clusters.

The values of GRB total energy ($Q$) and luminosity ($L$) close to real given in Table are ordinary for Galactic X-ray sources. For GRO J0422+32 $Q_{max}/Q_{min}$ = 18 whereas $L_{max}/L_{min}$ ~ 50.

Preliminary studies show that the GRB cluster signatures are observed from the most of X-NOVAE, listed in (White & van Paradijs 1996), as well as from Cyg X-1, which is belonged to high-mass BHXN and is persistent source of high-energy emission. It may be assumed from available information that BHXN in the quiescent state also emit single GRBs.

It becomes clear that the characteristic lifetime of the clusters (3-4 months) and their periodicity (4 years), show, respectively, the duration of minioutbursts in the light curves and the repeated manifestation of the BHXN activity.

## 3. Discussion and conclusions

The solution of the problem of cosmic GRB origin became possible due to discovery of the GRB clusters and their interpretation as a source image. Now we may state that considerable part the observed GRBs is a Galactic phenomenon and their source is a BH in the binary system. It is clear that the main difficulty in understanding GRB phenomenon was associated with BH participation in generation of GRBs: the large space-time curvature near the BH leads to the shift of GRB position relative to that of the source (BH). Discovery of GRB clusters reveals this effect (which is not pure gravitational lensing) and makes possible determination of the true position of the source and its identification.

The range and diversity of emissions observed from BHXN during its activity period, including GRBs produce quite a complicated picture. In the active phase of BHXN the accretion onto compact object, as assumed, manifest itself as rare outburst. During such an outburst an X-ray flux may increase in several days over several orders of magnitude and be accompanied by typical minioutbursts, optical and radiowave, in the light curves. Characteristics of X-NOVAE and manifestations of their activity are considered (Chen et al. 1997). The duration of emission in the active phase amounts to several months, the periodicity of activity is about 5 years. During their activity BHXN may expel superluminal jets. The clusters of gamma-ray bursts observed from BHXN just in the period of their activity copy with the certain discreteness the maxima of the light curves. The energy of photons in a GRB can reach up to 10 GeV (Hurley et al. 1994).

Obviously there exists one and same trigger for the generation of the radiation observed directly from the source (accretion disk?) and of GRB clusters with the afterglow shifted relative to the source position, but their separation depends on the distance to BH. In both cases, light curves are characterized by the fast rise and slow decay, but afterglow has a much shorter duration. By the order of magnitude the afterglow duration in X-rays is estimated as a day, while the X-ray light curve from X-NOVAE is being observed during many months. In the optical the distinction in duration is less considerable.

The origin of GRBs in our Galaxy, one of the main sources of which are X-NOVAE, permits a search BHXN from the singled out clusters and, on the contrary, to observe the GRB source, that is BH, over well known BHXN. Obviously the catalog of GRB sources (Kuznetsov 2000) is a catalog of BHXN. Observation of GRB, particularly, their clusters from the stellar objects, is a criterion of a BH discovery, its identification among other compact objects.

The physical dependence between GRBs and the afterglows on the one hand and outbursts in X-rays, optical and etc. from BHXN, on the other, may be understood due to simultaneous observations of BHXN in the period of activity and GRB clusters with the afterglows.

It should be mentioned that binary systems with a black hole as a primary star were considered as a probable source of GRBs (Brown et al, 2000); however, it was exactly the BH that hindered experimental determination of that relationship. Properties of GRB clusters, in particular, the absence of GRB without gravitational shift, constrain to believe that radiation of GRBs occur near the black hole under the assumption that the characteristic distance is of the order of the gravitational radius ($r_g$). There is no doubt that the mechanism of GRB generation by a BH will be updated in the nearest future.

If we know that the GRB source is a BH in the Galaxy, it is valid to state that a redshift ($z$) observed in the optical afterglow first of all represents a gravitational redshift which permits estimation of the gravity potential ($\varphi$) at the radiation point, $z_g = \varphi\, c^{-2}$ and other characteristics. Knowledge of the BH properties opens up new possibilities to explain old problematic phenomena, somehow related with black holes. We could mention among them SS 433, superluminal jets from BHXN GRS 1915+105 and GRO J1655-40. Here it is necessary to note that values range of redshifts, observed from quasars and GRBs (afterglow) are practically coincided.. There are the reviews of the most interesting observational data (Blandford & Gehrels 1999) about the black holes and their manifestations and theoretical concepts (Novikov & Frolov 2001).

Black holes in the binary systems form the real fireworks (see Fig. 1) in all spectrum of the electromagnetic radiation, beginning with GRBs (duration ~ min), which flash up in a cluster in turn with an interval about 1 week. Each GRB is accompanying by afterglow in X-rays (duration ~ up to a few days), in optics (duration ~ up to a few months), in radio wavelengths, etc. Evidently, the practical model of the GRB source may be named BH fireworks.

Gamma - ray bursts and afterglows are subject to the influence of large space-time curvature due to their proximity to a BH. Above the all durations ($\Delta t$) were measured by a distant observer. These phenomena, according to general relativity will proceed many orders of magnitude slower near the BH, relative to the local time scale, $\Delta \tau = (1 - r_g / r)^{0.5} \Delta t$ and will also show increasing redshift $z = (1 - r_g / r)^{-0.5} - 1$. At $r \to r_g$ both effects slowdown and redshift will tend to infinity. GRBs and afterglows emerge practically simultaneously, a delay may be of the order of tens of seconds. In the case of the gravitational collapse of gamma-radiation pulse onto the black hole a terrestrial observer would detect consecutive increase of wavelength (reddening) and of duration of fading electromagnetic radiation, which may be called afterglow. Apparently for Galactic gamma-ray bursts this is the most probable process, which gives a reasonable explanation of the observed afterglow all over the wavelength range. From above equations we may estimate the effects of slowdown and of wavelength increasing using the typical values of the GRB photon energy $E$ = 0.5 MeV and GRB duration $T$ = 1 min, X-ray and optical afterglow duration 100 min and $2 \cdot 10^5$ min, respectively. Then the slowdown coefficients $(1 - r_g / r)^{0.5}$ are equal $10^{-2}$ and $0.5 \cdot 10^{-5}$ determine, that X-ray and optical wavelengths will correspond to photon energy 5 keV and 2.5 eV. A distance relative to a gravitational radius for X-ray afterglow is defined according to $r_x = (1 + 10^{-4}) r_g$, and for optical - $r_{opt.} = (1 + 2.5 \cdot 10^{-11}) r_g$. Taking $r_g$ = 10 km, the distance above $r_g$ for X-rays is 1 m and for optical ~ $10^{-5}$ cm. Though it was evidently the given simplified estimation confirms once more that the afterglow goes from an event horizon of a black hole.

It is very important that the afterglow then may offer the unique possibility for study and testing of general relativity. By the way note that the value of true duration of GRB, measured over the local time scale, will be significantly lower that fixed by a terrestrial observer.

Nearly 30 years most important and topical problems in astrophysics have been search for black holes and origin of cosmic gamma-ray bursts. Discovery of GRB clusters and their interpretation as an image of the source forming space-time curvature, which evidently is typical for BH, lead to close relationship and to possibility of simultaneous solution of these problems.

Observation of space-time geometry near a BH (GRB clusters) and of redshift (z) effect confirming major consequences of gravity theory, eventually establishes the existence of black holes having stellar masses whose properties could be relatively easily studied. In fact, from this moment on, as we hope, the practical study and explanation of phenomena associated with black holes physics and testing of general relativity in the strong gravitational fields become possible.